\newcommand{\rem}[1]{}

\documentclass[prl,twocolumn,showpacs,preprintnumbers,amsmath,amssymb]{revtex4}
\usepackage{hyperref}
\usepackage{epsfig}
\usepackage{dcolumn}	% Align table columns on decimal point
\usepackage{bm}		% bold math
\begin{document}

\title{
Comment to ``Experimental Evidence of a Dynamic Jahn-Teller Effect in
C$_{60}^+$''
}

\author{Nicola Manini}
%\email{nicola.manini@mi.infm.it} % please insert if space is available
\affiliation{
Dipartimento di Fisica, Universit\`a di Milano,
Via Celoria 16, 20133 Milano, Italy  \\
and INFM, Unit\`a di Milano, Milano, Italy
\quad\quad{}
}
\author{Erio Tosatti}
%\email{tosatti@sissa.it} % please insert if space is available
\affiliation{
International School for Advanced Studies (SISSA),
Via Beirut 4, 34014 Trieste, Italy \\
INFM Democritos National Simulation Center, Trieste, Italy, and\\
International Centre for Theoretical Physics (ICTP), P.O. Box 586,
34014 Trieste, Italy
}

\date{\today}
\maketitle

A recent Letter %by Canton {\it et al.}\
\cite{Canton02etal} reports
photoemission (PE) data for the free C$_{60}$ molecule, showing an
interesting three-peak structure, presented as evidence of dynamic
Jahn-Teller (DJT) effect in the C$_{60}^+$ ion.
Those data constitute, along with earlier spectra by the Uppsala group
\cite{Bruhwiler97etal}, the best available piece of experimental evidence
about the spectrum of a hole in
%a free 
fullerene.
DJT must indeed affect the fivefold-degenerate $h_u$ hole molecular orbital
\cite{Manini01etal},
%and any spectroscopic evidence thereof is very welcome,
but we contend that the energy separation of these three peaks is far too
large for the proposed tunneling interpretation to be correct.

In detail, the observed structure is claimed to indicate a $D_{3d}$
distortion in the lowest-energy JT well, accompanied by tunneling
between the local minima.
The main argument offered is that the ten $D_{3d}$ valleys 
yield three tunnel-split states $H_u$\,(5) + $G_u$\,(4) +
$A_u$\,(1), to be identified with the three observed structures.
Alternatively, six $D_{5d}$ valleys would split into two $H_u$ (5) +
$A_u$\,(1), while 
valleys of lower symmetry ($D_{2h}$,
$C_{2h}$) would split more than threefold. 
Given this interpretation however, the tunnel splitting magnitudes
between $D_{3d}$ valleys would amount to 230~meV ($H_u-G_u$), and
390~meV ($H_u-A_u$).  Both splittings are alarmingly larger than any of the
vibrational frequencies of fullerene ($32\div 195$~meV).

\begin{figure}[b]
\epsfig{file=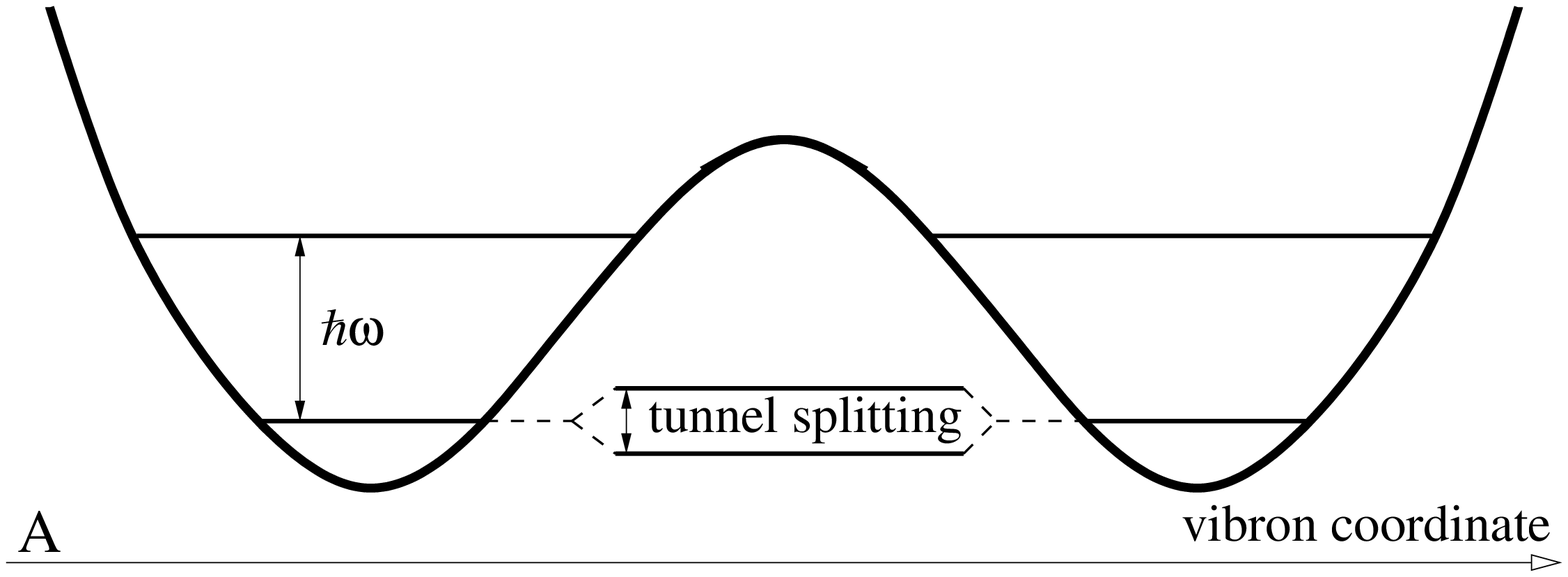,width=79mm}
\epsfig{file=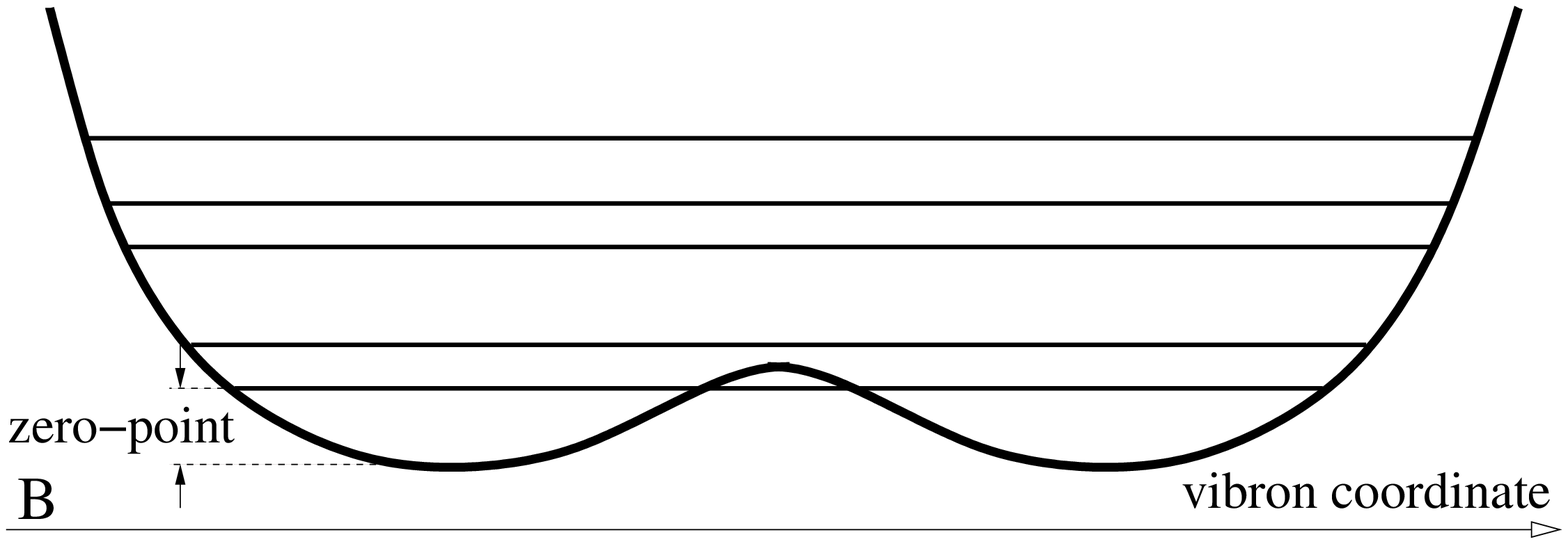,width=79mm}
\caption{\label{doublewell:fig}
%A scheme of the two limiting cases of weak (A) and strong (B) tunneling.
A pictorial of (A) tunneling  and (B) extended anharmonic excitations.
%In (B) the latter, all the levels in the wells are intermixed.
}
\end{figure}

Our contention is precisely that these supposed splittings are far too
large for the tunneling interpretation to be valid.
Tunnel splittings only make sense in the large-barrier limit, when 
they are smaller than the {\em smallest} vibrational
quantum $\hbar \omega$, here of 32~meV (Fig.~\ref{doublewell:fig}A), 
and the model of cited Ref.~[36] applies.
In the opposite limit (Fig.~\ref{doublewell:fig}B), barriers
between valleys are lower than the kinetic energy, and tunneling-split
levels are replaced by extended anharmonic excitations,
delocalized over all valleys.
To estimate the two lowest excitations to be expected for
C$_{60}^+$, we %have 
carried out a realistic calculation of the lowest 
$H_u$, $G_u$, and $A_u$ vibronic states of % C$_{60}^+$.
%
%Our Hamiltonian consists of
an %fivefold
$h_u$ electronic level JT
coupled to eight fivefold $H_g$ and to six fourfold $G_g$ vibrational modes
\cite{Manini01etal}.
Using the {\it ab-initio} JT coupling parameters and frequencies of 
Ref.~\cite{Manini01etal} in a symmetry-restricted Lanczos 
diagonalization, we obtain 18~meV and 30~meV ($\lesssim\hbar \omega$) for
the excitation energies from the $H_u$ ground state to the lowest $A_u$ and
$G_u$ vibronic states respectively.
These values are % clearly
one order of magnitude smaller than
% those they should have according to the interpretation of 
the PE structures \cite{Canton02etal}.

Is this discrepancy due to uncertainty in the precise values
of the coupling parameters? We think not.
If the actual couplings were smaller or, as is more likely, larger than the
those assumed, or if they made $D_{3d}$ wells lower than $D_{5d}$
wells, then the lowest $A_u$, $G_u$, and $H_u$ states would always lie within
a range of $\sim\hbar\omega \simeq 30$~meV.
%
% CHANGE
The spectral structures above 200~meV might reflect high frequency 
vibrons, or else they might be due to an
electronic splitting, in analogy to the interpretation of the PES of
Fe(CO)$_5$ in cited Ref.~[23].
%
%Accordingly, CHANGE
The observed splittings would be in the same range ($\approx
180$~meV) as the computed electronic JT splitting of Fig.~3 of
Ref.~\cite{Manini01etal}.
A full calculation of the spectrum that will include both low-energy
(tunneling) and high-energy ``electronic'' splittings could in principle be
done based on Fermi's golden rule \cite{Martinelli91etal}, but is at
present still unavailable.
%CHANGE

In summary,
%while the PE data of C$_{60}$ must
%certainly be heavily affected by dynamic JT in the final state, 
it would seem that the three-peak
structure, though certainly related to DJT,  
does not particularly provide evidence for tunneling among
$D_{3d}$ valleys as claimed.
%
%A detailed %explanation assignment  
A full explanation of this spectrum must %therefore 
await further work.

%%%%%%%%%%%%%%%%%%%%%%%%%%%% REFERENCES %%%%%%%%%%%%%%%%%%%%%%%%%%%%%%%%%%%%

%\bibliographystyle{apsrev}
%\bibliography{biblio}

\end{document}